\documentclass[12pt]{article} 
\usepackage{amsmath}
\usepackage{amsfonts}
\usepackage[final]{graphicx}
\usepackage{times}
\usepackage{latexsym}
\begin{document}
\begin{center}
\textbf{{\Large Bounds for codes  for a non-symmetric ternary channel}} \\
\vspace{5mm}
Ludo Tolhuizen, Philips Research Laboratories, Eindhoven, The Netherlands; \\
\verb+ ludo.tolhuizen@philips.com+ 
\end{center}
\newtheorem{theo}{Theorem}
\section{Introduction}
In \cite{ternary1}, a
non-symmetric ternary communication channel inspired by 3-valued
semiconductor memories was introduced, and error-correcting coding for this channel was studied.
The authors of \cite{ternary1}
showed the relevance of the minimum $d_1$-distance (defined below)
of a ternary code for judging its error-correcting capabilities on
this channel, gave a code construction, and derived a Hamming-like upper
bound on the size of a code of given length and minimum $d_1$-distance. 
The work was extended in \cite{ternary}, where the authors obtained the channel capacity,
and constructed optimal codes with a short length 
by techniques for finding
cliques in graphs.

In the present paper, we give upper and lower bounds on the size of
codes for the $d_1$-distance. We first introduce some notation.

 We consider codes over the ternary alphabet
$Q=\{-1,0,1\}$. For ${\mathbf x},{\mathbf y} \in Q^n$, we define $d_1({\bf
x},{\bf y})$ as
\[ d_1({\bf x},{\bf y}) = \sum_{i=1}^n  |x_i - y_i|. \]
For each $C\subseteq Q^n$, we denote the minimum
$d_1$-distance between any two different words of $C$ by $d_1(C)$.
Furthermore, we define
\[ T(n,d) = \max \{ |C| \mid C\subset Q^n \mbox{ and }
d_1(C)\geq d\}. \] It is our aim to provide upper and lower bounds
on $T(n,d)$. In the remainder of the paper, when we speak about
"distance", we mean $d_1$-distance.

Unlike the Hamming
distance, the $d_1$-distance is {\em not} translation-invariant. For example,
the number of words at distance one from the all-zero word
of length $n$ equals $2n$, while the number of words of distance one
from the all-one word of length $n$ equals $n$. As a result,
many bounds for codes in Hamming space \cite{McWSl} do not readily
translate to codes for the $d_1$-distance. The Hamming bound from
\cite{ternary1}, for example, takes into account the largest balls
and hence seems to be rather weak.

Some code constructions and bounds will use results for codes for
the Hamming metric. The Hamming distance between two vectors ${\bf x}$ and ${\bf y}$ of equal length
is denoted as $d_{\rm H}({\bf x},{\bf y})$. The minimum
Hamming distance of a code $C$ is denoted as $d_{\rm H}(C)$, and we define
\[ A_q(n,d) = \max\{ |C| \mid C\subset\{0,1,\ldots q-1\}^n \mbox{
and } d_{\rm H}(C)\geq d\}. \]
\section{Bounds from code shortening and puncturing}
Let $C\subseteq Q^n$ have minimum $d_1$-distance $d$.
For $i\in Q$, we define
\[ C_i = \{ (x_1,x_2,\ldots x_{n-1}) \mid (x_1,x_2,\ldots
,x_{n-1},i)\in C \} . \] We have the following easy proposition.
\newtheorem{prop}{Proposition}
\begin{prop}\label{prop:shorten}
For each $i\in Q$, we have that $d_1(C_i) \geq d$, so $|C_i|\leq T(n-1,d)$; \\
 moreover, we have that $d_1(C_0\cup C_1)\geq d-1$, so $|C_0|+ |C_1|\leq T(n-1,d-1)$; \\
 and finally, $d_1(C_0\cup C_1\cup C_{-1}) \geq d-2$.
 \end{prop}
 \newtheorem{cor}{Corollary}
 \begin{cor} The following inequalities are valid:
% \begin{eqnarray}
\begin{align}
 T(n,d) & \leq  3 T(n-1,d) \label{puncture} \\
 T(n,d) & \leq T(n-1,d)+ T(n-1,d-1) \label{mix} \\
 T(n,d) & \leq T(n-1,d-2) \label{shorten}
% \end{eqnarray}
\end{align}
 \end{cor}
{\bf Proof.}
Let $C$ have length $n$, minimum distance $d$, and size $T(n,d)$. \\
Inequality~(\ref{puncture}) follows from the fact that
$T(n,d)= |C| = |C_{-1}| + |C_0| + |C_1| \leq 3 T(n-1,d)$,
where the inequality follows from the first statement of Proposition~\ref{prop:shorten}.
\\ Inequality~(\ref{mix}) follows from the fact that $T(n,d) = |C| = |C_{-1}| + |C_0\cup C_1| \leq T(n-1,d) + T(n-1,d-1)$,
where the inequality follows from the two first statements of Proposition~\ref{prop:shorten}.
\\ Inequality~(\ref{shorten}) is a direct consequence of the final statement in Propostion~\ref{prop:shorten}.
$\;\;\;\;\Box$
            \begin{cor} For $n\geq 1$, we have that
            $T(n,2)=(3^n+1)/2$.
            \end{cor}
{\bf Proof.}
Using Inequality~(\ref{mix}) and
induction on $n$, one readily finds that
$T(n,2)\leq\frac{1}{2}(3^n+1)$. The code
consisting of all vectors of length $n$ containing an even number of
zeros, which is a special case of the construction in \cite{ternary1},
 has minimum distance two and $\frac{1}{2}(3^n+1)$ words.$\;\;\;\;\Box$
 
\section{Bounds and constructions based on codes for the \\ Hamming distance}
For any two ternary vectors ${\bf
x}$ and ${\bf y}$ of equal length, we clearly have that
$d_{\rm H}({\bf x},{\bf y})\leq d_1({\bf x},{\bf y}) \leq 2 d_{\rm H}({\bf
x},{\bf y}).$ As a consequence, for any ternary code $C$, we have
that
$d_{\rm H}(C) \leq d_1(C) \leq 2d_{\rm H}(C)$, and so
\begin{equation}\label{ternarybound}
 A_3(n,d) \leq T(n,d) \leq A_3\left(n,\left\lceil\frac{d}{2}\right\rceil\right).
 \end{equation}
Also, if ${\bf x}$ and ${\bf y}$ are two vectors over $\{-1,1\}$,
then $d_1({\bf x},{\bf y})=2d_{\rm H}({\bf x},{\bf y})$, and so
\begin{equation}\label{binlow}
 T(n,d) \geq A_2\left(n,\left\lceil\frac{d}{2} \right\rceil \right).
 \end{equation}
 \begin{prop}\label{prop:2}
 We have that $\left( \frac{3}{4}\right)^n \cdot A_2(2n,d) \leq T(n,d)\leq A_2(2n,d)$.
 \end{prop}
 {\bf Proof.} 
 We define the mapping $\phi:Q\mapsto A:=\{(0,1),(0,0),(1,0)\}$ as
\[ \phi(-1)=(0,1), \phi(0)=(0,0), \mbox{ and } \phi(1)=(1,0), \]
and extend it to a mapping from $Q^n$ to $A^n$ by applying
$\phi$ component-wise. \\ It is clear that for any ${\bf x}$ and ${\bf
y}$ in $Q^n$, we have 
$d_{1}({\bf x},{\bf y}) = d_{\rm H}(\phi({\bf x}),\phi({\bf y}))$. 
As a consequence, for each $C\subseteq Q^n$, we have $d_1(C) =
d_{\rm H}({\phi(C)})$, which implies the upper bound on $T(n,d)$. \\
Conversely, let $C\subset \{0,1\}^{2n}$ have minimum
Hamming distance $d$. For each ${\bf c} \in C$, there are $|A|^n$ vectors ${\bf x}$ such that ${\bf x}+{
\bf c} \in A^n$. Hence, for at least one of the $2^{2n}$ choices for ${\bf x}$, the size of 
$({\bf x}+ C)\cap A^n$ is at least $|C|A^n/2^{2n}$. As the minimum $d_1$-distance of
$\phi^{-1}(({\bf x}+C) \cap A^n)$ equals $d_{\rm H}(({\bf x}+C)\cap A^n)\geq d_{\rm H}({\bf x}+C)=d$,
the lower bound follows.
$\;\;\;\Box$
\\ \\
The elegant construction from \cite{ternary1} yields the following theorem.
\begin{theo}\label{theo:constr}
Let $C$ be a binary code with minimum Hamming distance $d$ and $A_w$
words of Hamming weight $w$ ($w=0,1,\ldots ,n$). Then
\[ T(n,d) \geq \sum_{w=0}^n A_wA_2\left(w,\left\lceil\frac{d}{2}\right\rceil\right). \]
\end{theo}
By averaging Theorem~\ref{theo:constr} over all cosets of $C$, we obtain the
following corollary. \begin{cor}\label{constrav} 
\[T(n,d) \geq
\frac{A_2(n,d)}{2^n} \sum_{w=0}^n {n\choose w}
A_2\left(w,\left\lceil\frac{d}{2} \right\rceil \right) . \]
\end{cor}
\section{Plotkin bound}
Theorem~\ref{th:Plotkin} below is an anlogon to the Plotkin bound for codes in Hamming space
\cite[Sec.\ 2.2]{McWSl}, and is proved in Appendix~A.
\begin{theo}\label{th:Plotkin}
For $d>n$, we have that $T(n,d)\leq \frac{d}{d-n}$. A code attaining equality is a code over $\{-1,1\}^n$ with minimum Hamming distance $d/2$ satisfying the binary Plotkin bound. \\
Moreover, we have that
\[ T(d,d) \leq 2d + \frac{1}{2} + \sqrt{2d+\frac{1}{4}}\; .\]
\end{theo}
 
\section{Gilbert-Varshamov bounds}
In this section, we derive lower bounds on $T(n,d)$ using the same arguments as for the Gilbert-Varshamov (GV) bound in Hamming space. The GV bound for codes with the Hamming
metric guarantees the existence of a $q$-ary code of length $n$ and
minimum Hamming distance $d$ with a cardinality at least $q^n/V_q(n,d-1)$,
where $V_q(n,r)$ denotes the cardinality of a ball of radius $r$ in
$\{0,1,\ldots ,q-1\}$.
The volume of a ball in the $d_1$-metric
depends on its center. The generalized GV bound \cite{GV}
guarantees the existence of a code of length $n$ and minimum
$d_1$-distance $d$ with cardinality at least $3^n/\bar{V}(n,d-1)$,
where $\bar{V}(n,d-1)$ is the {\em average} size of a ball of
radius $d-1$ in $Q^n$ endowed with the $d_1$-metric. 
For computing this average size, we define $m(n,w)$ to be the number of ordered pairs of vectors in $Q^n$
that have $d_1$-distance $w$. By induction on $n$, one readily obtains the following proposition.
\begin{prop}
We have that $\sum_{w=0}^{2n} m(n,w)z^w = (3+4z+2z^2)^n$.
\end{prop}
By writing $(3+4z+2z^2)^n=(2(1+z)^2+1)^n$, expanding using the
binomial theorem, and collecting terms of equal power, we obtain that
$ m(n,w) = \sum_{i=0}^n {n \choose i}2^i {2i \choose w}.$
The generalized GV bound \cite{GV} thus implies that
\begin{equation} \label{eq:ggv}
T(n,d) 
\geq \frac{3^{2n}}{\sum_{w=0}^{d-1} m(n,w)} =
\frac{3^{2n}}{\sum_{w=0}^{d-1} \sum_{i=0}^n {n \choose i} 2^i {2i \choose w}}.
\end{equation}
%\subsection{Constant-weight ternary codes}

By applying the GV-argument to $Q^n_w$, the set of words in $Q^n$ of Hamming weight $w$, we obtain that there exists a code in $Q^n_w$ with minimum distance $d$ and cardinality at least $|Q^n_w|/ V(n,d-1,w)$, where $V(n,d-1,w)$ is the number of words in $Q^n_w$ at distance at most $d-1$ from a fixed word in $Q^n_w$. It is clear that $|Q^n_w|={n\choose w}2^w$. For obtaining $V(n,d-1,w)$,
we use the following proposition. 
\begin{prop} For each ${\bf x}\in Q^n_w$ and each integer $i$, we have that
\[ |\{{\bf y} \in Q^n_w\mid d_1({\bf x},{\bf y})=2i+\epsilon\}| =
\left\{ \begin{array}{ll} 0 & \mbox{ if } \epsilon =1, \\
                  \sum_j{w\choose j}{w-j\choose i-j}{n-w\choose j}2^j & \mbox{ if } \epsilon = 0. 
                   \end{array} \right. \]
\end{prop}
{\bf Proof.}                   
Let ${\bf x}\in Q^n$ start with $w$ ones and end in $n-w$ zeros.
Let ${\bf y}\in Q^n_w$. We define $j$ as the number of zeros in the $w$ leftmost positions in ${\bf y}$, and $w-i$ as the number of ones in the leftmost positions of ${\bf y}$. Then $(i-j)$ of the $w$ leftmost entries of ${\bf y}$ equal -1, and, as ${\bf y}$ has weight $w$, $j$ entries of the righmost values of ${\bf y}$ are non-zero. We conclude that the number of vectors ${\bf y}$ satisfying the above constraint equals
\[ {w\choose j} {w-j\choose i-j} {n-w \choose j}2^j , \]
while $d_1({\bf x},{\bf y})= j + 2(i-j)+ j = 2i. \;\;\;\;\Box$ 
\\ \\
We conclude that the following theorem holds.
\begin{theo}\label{th:constantweight}
For each $w, 1\leq w\leq n$, we have that 
\[ T(n,d) \geq \frac{{n \choose w}2^w}{\sum_{i=0}^{(d-1)/2} \sum_{j=0}^{\min(i,n-w,w)} {w \choose j} {n-w \choose j} {w-j \choose i-j} 2^j}. \]
\end{theo}
\section{Asymptotics of the bounds}
%\subsection{Finite length}
%In the full paper, we will give a table of the numerical results of
%the bounds for small parameter values.
In this section, we derive the asymptotic versions of the obtained bounds.
For $0 < \delta < 2$, we define
\[ \tau(\delta) = \lim_{n \to \infty} \sup \frac{1}{n} \log_3 (T(n,\lceil \delta n \rceil)) .\]
For $0 < \delta < 1 $, we define
\[ \alpha_q(\delta) = \lim_{n \to \infty} \sup \frac{1}{n} \log_q (A_q(n,\lceil
\delta n\rceil)) . \] 
We will use the asymptotic GV bound:
for $0\leq\delta\leq 1-\frac{1}{q}$, we have
\begin{equation} \label{GVHamming}
 \alpha_q(\delta) \geq 1 - h_q(\delta), 
 \end{equation}
where $h_q$ is the $q$-ary entropy function, defined as 
\begin{equation} \label{hdef}
 h_q(x) = -x\log_q(x) - (1-x)\log_q(1-x) + x\log_q(q-1) .
 \end{equation}
The following inequalities are readily
obtained from 
(\ref{ternarybound}), (\ref{binlow}), and
Proposition~\ref{prop:2}
\begin{align}
\alpha_3(\delta) &\leq \tau(\delta) \leq \alpha_3\left(\frac{\delta}{2}\right) \label{ternaryas} \\
 \tau(\delta) &\geq \log_3 (2) \alpha_2\left(\frac{\delta}{2}\right) \label{binlowas} \\
\tau(\delta) &\leq 2\log_3(2) \alpha_2\left(\frac{\delta}{2}\right) \label{binupas} \\
\tau(\delta) &\geq \log_3\left(\frac{3}{4} \right) +2\log_3(2)\alpha_2\left(\frac{\delta}{2}\right) \label{bindoubleas}
\end{align}
As $\alpha_2(\delta) > 0$ if and only if $\delta < \frac{1}{2}$, we
derive from (\ref{binlowas}) and (\ref{binupas}) that $\tau(\delta)
> 0$ if and only if $\delta <  1$.

The asymptotic version of Corollary~\ref{constrav} is
\begin{equation} \label{eq:cor31}
%\tau(\delta) \geq \log_3(2)\left(-1+\alpha_2(\delta)+\sup_{\substack{0 < \omega < 1 \\ \delta < \beta < 1}} \left \{ \mathbb{H}_2(\omega)+\omega \left(1-\mathbb{H}_2(\beta/2) \right) \right\} \right)
\tau(\delta) \geq \log_3(2)\left(-1+\alpha_2(\delta)+\sup_{\max(\delta,1/2) < \omega < 1} \left \{ h_2(\omega)+\omega \alpha_2(\delta/(2 \omega)) \right\} \right)\; .
\end{equation}
Using the asymptotic binary GV bound (\ref{GVHamming}) we find that (\ref{eq:cor31}) implies that
\begin{equation} \label{eq:cor3}
%\tau(\delta) \geq \log_3(2)\left(-1+\alpha_2(\delta)+\sup_{\substack{0 < \omega < 1 \\ \delta < \beta < 1}} \left \{ \mathbb{H}_2(\omega)+\omega \left(1-\mathbb{H}_2(\beta/2) \right) \right\} \right)
\tau(\delta) \geq \log_3(2)\left(-1+\alpha_2(\delta)+\sup_{\max(\delta,1/2) < \omega < 1} \left \{ h_2(\omega)+\omega \left(1-h_2(\delta/(2 \omega))\right) \right\} \right)\; .
\end{equation}
If $\delta\leq 1/2$, the supremum in (\ref{eq:cor3})  is attained for
%$\beta=\delta$ and $\omega=1/(1+2^{\mathbb{H}_2(\delta/2)-1})$. 
$\omega = \frac{1}{6}(2+\delta+\sqrt{4-8\delta+\delta^2})$.

The asymptotic form of the generalized GV bound (\ref{eq:ggv}) is
\begin{equation} \label{eq:ggv_ternary}
\tau(\delta) \geq 2- \log_3(2)\sup_{\substack{0 < \omega < 1\\ 0 < \beta <
\min(\delta,2\omega)}} \left \{h_2(\omega)+2\omega h_2\left(\frac{\beta}{2\omega} \right)+\omega
\right\}.
\end{equation}
If $\delta \geq 8/9$, the supremum in (\ref{eq:ggv_ternary}) is attained for $\omega=\beta=8/9$, and we obtain the trivial inequality $\tau(\delta)\geq 0$. \\ Otherwise, if $\delta < 8/9$,   we obtain, by setting the partial deriviatives with respect to $\beta$ and $\omega$ equal to zero, that the supremum in (\ref{eq:ggv_ternary}) is attained for
 $\beta=\delta$ and $\omega = (2+\delta+\sqrt{2(-\delta^2+2\delta+2)})/6$. 
%When $\delta < 8/9$, we get
%\begin{equation} \label{eq:ggv_ternary}
%\begin{split}
%\tau(\delta) &\geq 2- \mathbb{H}_3\left(\frac{2+\delta+\sqrt{2(-\delta^2+2\delta+2)}}{6} \right) \\
%&- \frac{2+\delta+\sqrt{2(-\delta^2+2\delta+2)}}{6} \left( 2
%\mathbb{H}_3\left(\frac{3 \delta}{2+\delta+\sqrt{2(-\delta^2+2\delta+2)}  } \right)+ \log_3(2) \right).
%\end{split}
%\end{equation}

The asymptotic version of Theorem~\ref{th:constantweight}, the GV bound for constant-weight
ternary codes, is the following.
For every $\omega, 0 < \omega < 1$, we have that
\begin{equation} \label{eq:asymptconstweight}
\begin{split}
\tau(\delta) \geq  \log_3(2) [
h_2(\omega)+\omega -
\sup_{\substack{0 < \beta < \min(\delta/2,\omega)\\0 < \gamma < \min(\beta,\omega,1-\omega)}} & 
\left \{ \omega h_2\left(\frac{\gamma}{\omega} \right)+(1-\omega) h_2\left(\frac{\gamma}{1-\omega} \right) \right.   \\
&\;\;\;+   \left. \left.(\omega-\gamma) h_2\left( \frac{\beta-\gamma}
{\omega-\gamma} \right)+\gamma \right\} \right] \; .
\end{split}
\end{equation}
If $\delta\geq \omega(2-\omega)$,  the supremum in (\ref{eq:asymptconstweight}) 
is attained for  
$\beta = \frac{1}{2}\omega(2-\omega)$ and $\gamma=\omega(1-\omega)$, and we 
obtain the trivial inequality $\tau(\delta)\geq 0$.  \\
If $\delta < \omega(2-\omega)$, the supremum in (\ref{eq:asymptconstweight}) is attained for
$\beta=\delta/2$ and $\gamma=1-\omega+\delta/2-\sqrt{(1-\omega)^2+\frac{1}{4}\delta^2}$. 
\\ Next, for fixed $\delta$, we optimize (\ref{eq:asymptconstweight}) over $\omega$, 
using the values for $\beta$ and $\gamma$ obtained before. As shown in Appendix~B, the optimzing value for $\omega$ equals
$\omega = \frac{1}{3}(1+\delta+\sqrt{\delta^2-\delta+1})$. \\
Note that for $\delta=0$, the optimizing $\omega$, as expected, equals $\frac{2}{3}$, while for $\delta=1$, the optimal value is $\omega=1$, i.e., for large $\delta$, binary codes are good.
\

%the right hand side of (\ref{eq:asymptconstweight}) is zero. When $\delta \leq \omega(1-\omega/2)$, (\ref{eq:asymptconstweight}) reduces to
%\begin{equation} \label{eq:asymptconstweight1}
%\tau(2\delta,\omega) \geq -\omega \mathbb{H}_3(1-\omega)+\omega \mathbb{H}_3(\omega)
%\end{equation}

\section{Comparison of asymptotic lower bounds}
\begin{figure}[tbp]
 \centerline{\includegraphics[width=\columnwidth]{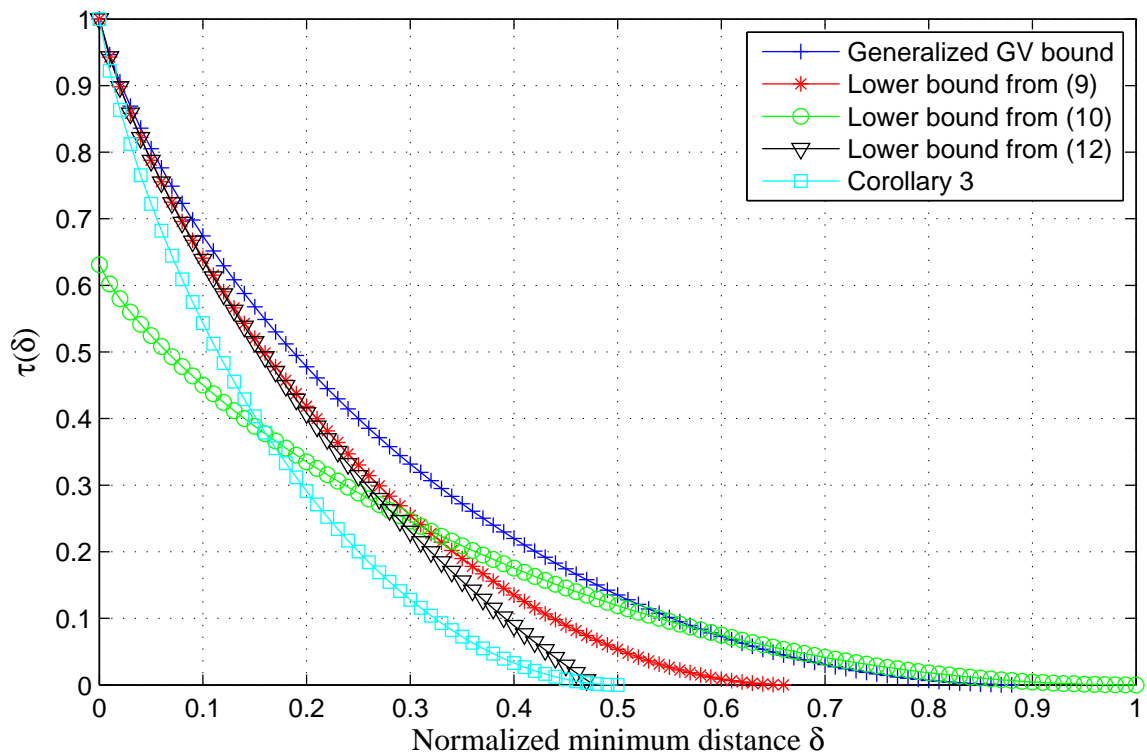}}
 %\vspace{-3mm}
   \caption{Lower bounds on $\tau(\delta)$.}
    \vspace{-3mm}
   \label{fig:1}
\end{figure}

In Figure~\ref{fig:1}, we plot the various asymptotic lower bounds on
$\tau(\delta)$.  %The optimization problem for the generalized
%Gilbert-Varshamov bound has been solved numerically.
%
We have used the GV bound (\ref{GVHamming}) to lower bound $\alpha_q(\delta)$.
It is interesting to see that for large $\delta$, the bound from
(\ref{binlowas}), obtained using the GV bound for binary codes,
performs better than the generalized GV bound for
ternary codes with the $d_1$-distance from (\ref{eq:ggv_ternary}). This shows some similarity to
the result from \cite{ternary} that states that for large cross-over
probabilities, channel capacity is achieved by a binary code.
\begin{figure}[tbp]
 \centerline{\includegraphics[width=\columnwidth]{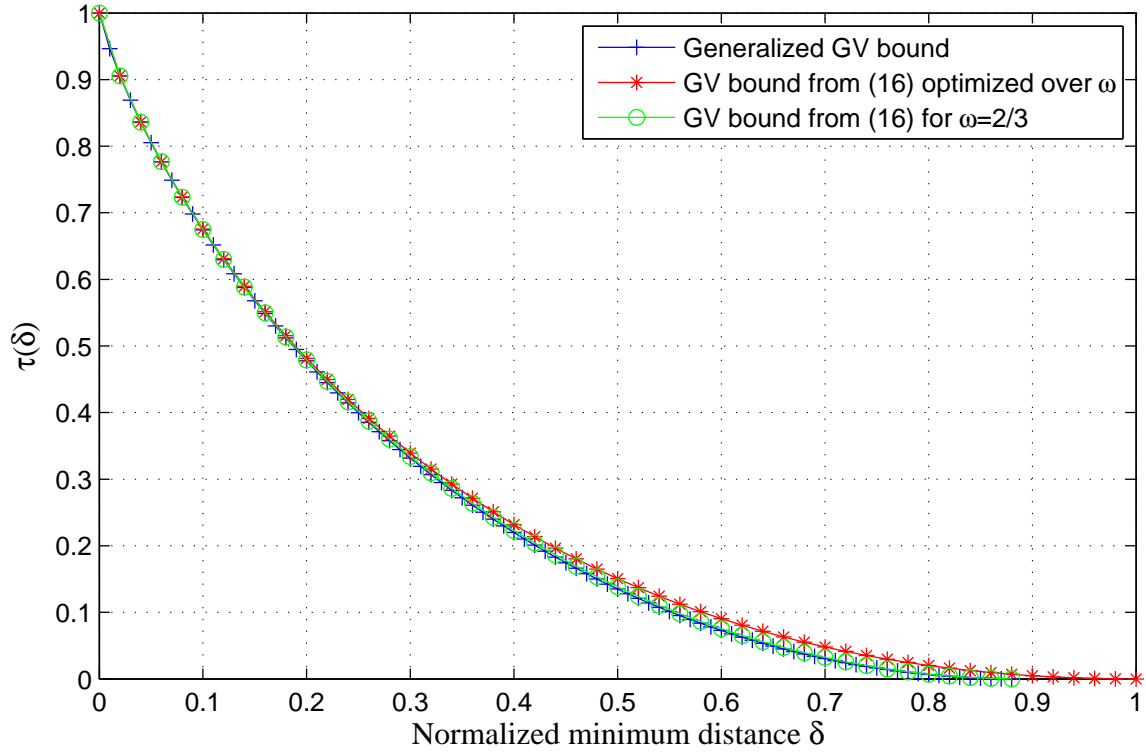}}
 %\vspace{-3mm}
   \caption{Lower bounds on $\tau(\delta)$.}
    \vspace{-3mm}
   \label{fig:2}
\end{figure} \\
In Figure~\ref{fig:2}, we plot the generalized GV bound and the bound from (\ref{eq:asymptconstweight}) optimized over $\omega$, i.e., for each value of $\delta$ the expression from (\ref{eq:asymptconstweight}) is optimized over $\omega$. For comparison purposes, we have also plotted the bound from (\ref{eq:asymptconstweight}) for $\omega=2/3$. Note that the asymptotic GV bound for constant-weight codes (\ref{eq:asymptconstweight}) slighly improves the generalized GV bound when we optimize over $\omega$.

\section*{Acknowledgement}
The author wishes to thank Eirik Rosnes (Selmer Center, Dept.\ of Informatics, University of Bergen, Norway) for his help
with the calculus for the asymptotic GV bounds and for producing the figures.

\section*{Appendix A: Proof of Theorem~\ref{th:Plotkin}}
Let $C$ be a code of length $n$, minimum distance $d$, and $M$ words.
For $1\leq i\leq n$ and $j\in Q$, we define 
$m_j(i) = | \{ {\bf c}\in C \mid c_i=j \} |$.
We define $S$ as 
$ S = \sum_{{\bf x}\in C}\sum_{{\bf y}\in C} d_1({\bf x},{\bf y})$. 
We obviously have that $S\geq M(M-1)d$. On the other hand, 
\begin{equation} \notag
\begin{split}
S &= \sum_{i=1}^n m_0(i)[m_{1}(i)+m_{-1}(i)] + m_1(i)[m_0(i)+2m_{-1}(i)]
                       + m_{-1}(i)[m_0(i)+2m_{1}(i)] \\
&= \sum_{i=1}^n 2m_0(i)[m_1(i)+m_{-1}(i)] + 4m_{1}(i)m_{-1}(i).
\end{split}
\end{equation}
 For each $i\in\{1,\ldots ,n\}$, we have that
 \[ 4m_{1}(i)m_{-1}(i) = (m_{1}(i)+m_{-1}(i))^2 - (m_1(i)-m_{-1}(i))^2 \leq 
                 (m_1(i)+m_{-1}(i))^2. \]
Moreover, as $m_1(i)+m_{-1}(i)=M-m_{0}(i)$, we obtain that
\begin{equation} \notag
\begin{split}
S \leq \sum_{i=1}^n 2m_0(i)(M-m_0(i))+(M-m_0(i))^2 &=
        \sum_{i=1}^n (M-m_0(i))(M+m_0(i)) \\
&= nM^2 - \sum_{i=1}^n m_0^2(i) .
\end{split}
\end{equation}
As a consequence, we have that 
\begin{equation}\label{eq:ternaryplotkin}        
M(M-1)d\leq S \leq nM^2 - \sum_{i=1}^n m_0^2(i).
\end{equation}
Inequality~(\ref{eq:ternaryplotkin}) clearly implies that $M(M-1)d\leq nM^2$, and so, if
        $d>n$, we have that $M\leq d/(d-n)$. If equality holds, we must have that $\sum_{i=1}^n m_0^2(i)=0$, so  $C\subset \{-1,1\}^n$. \\
If $d=n$, (\ref{eq:ternaryplotkin}) says that
  $\sum_{i=1}^n m_0^2(i) \leq Mn$.
  Hence, for some $i$, we have that $m_0^2(i)\leq M$.    
  By shortening in position $i$,w e obtain three codes of length $n-1$ and minimum distance $d$, and
  (using the same notation as in Section 2), we find that
 \[ M = |C_0| + |C_{-1}| + |C_1| \leq \sqrt{M} + 2T(d-1,d) \leq \sqrt{M} + 2d , \mbox{ whence }
 (\sqrt{M}-\frac{1}{2})^2 \leq 2d+\frac{1}{4}. \;\;\;\Box \]    

\section*{Appendix~B: Optimizing (\ref{eq:asymptconstweight}) over $\omega$}
Setting the partial derivative of the right hand side of (\ref{eq:asymptconstweight}) with respect
to $\omega$ to zero, with $\beta=\delta/2$ and $\gamma=1-\omega+\delta/2-\sqrt{(1-\omega)^2+\frac{1}{4}\delta^2}$,  results in the following equation:
\begin{equation} \label{eq:optimalomega}
%\left( \frac{\gamma^2}{2(\delta-\gamma)(1-\omega-\gamma)} \right)^{\gamma'} \left( \frac{1-\omega}{\omega} \right)^2 \frac{2(\omega-\delta)}{1-\omega-\gamma} = 1
3 \omega^4-2(4+\delta) \omega^3+4(1+2\delta)\omega^2-2\delta(2+\delta)\omega+\delta^2=0.
\end{equation}
The polynomial in (\ref{eq:optimalomega}) factors as $(\omega^2-2\omega+\delta)(3\omega^2-2(\delta+1)\omega+\delta)$,
so that we find the four roots for the polynomial, viz.
\[ 1-\sqrt{1-\delta},\; 1+\sqrt{1-\delta},\; \frac{1}{3}(1+\delta-\sqrt{\delta^2-\delta+1}),
\mbox{ and } \frac{1}{3}(1+\delta+\sqrt{\delta^2-\delta+1}).\]
%Using Cardano's formula for solving a general fourth order polynomial equation, results in the following solutions:
%\begin{displaymath}
%\omega = \begin{cases}
%1+\sqrt{1-2\delta} \\
%1-\sqrt{1-2\delta} \\
%\frac{1}{3}\left(2\delta+1+\sqrt{4\delta^2-2\delta+1} \right) \\
%\frac{1}{3}\left(2\delta+1-\sqrt{4\delta^2-2\delta+1} \right) \end{cases}
%\end{displaymath}
Note that in order to find a non-zero solution in (\ref{eq:asymptconstweight}), it is required that $\delta >  \omega(2-\omega)$, and $\frac{1}{3}(1+\delta+\sqrt{\delta^2-\delta+1})$ is the only root satisfying this requirement.
%The only solution within the range $(1-\sqrt{1-\delta},1)$ is $\omega = \frac{1}{3}\left(\delta+1+\sqrt{\delta^2-\delta+1} \right)$. Note that $\frac{1}{3}\left(\delta+1-\sqrt{\delta^2-\delta+1} \right)$ is always smaller than $1-\sqrt{1-\delta}$.
%%
%where $\gamma' = -1+\frac{1-\omega}{\sqrt{(1-\omega)^2+\delta^2}}$ is the derivative of $\gamma$ with respect to $\omega$.
%Thus, an optimal value for $\omega$ is a solution to (\ref{eq:optimalomega}) in the range $[1-\sqrt{1-2\delta},1]$.

         \end{document}